# Nitrogen Loss from Pluto's Birth to the Present Day via Atmospheric Escape, Photochemical Destruction, and Impact Erosion

Perianne E. Johnson[1,2], Leslie A. Young[3], David Nesvorný[3], and Xi Zhang[4]
[1] University of Colorado, Boulder, CO, USA; perianne.johnson@jsg.utexas.edu
[2] University of Texas Institute for Geophysics, Austin, TX, USA
[3] Southwest Research Institute, Boulder, CO, USA
[4] University of California, Santa Cruz, CA, USA

## Abstract

We estimate the loss of nitrogen from Pluto over its lifetime, including the giant planet instability period, which we term the "Wild Years." We analyze the orbital migration of 53 simulated Plutinos, which are Kuiper Belt Objects (KBOs) captured into 3:2 mean-motion resonance with Neptune during the instability. This orbital migration brought the Plutinos from 20 to 30 au to their present-day orbits near 40 au along a nonlinear path that includes orbits with semimajor axes from 10 to 100 au. We model the thermal history that results from this migration and estimate the volatile loss rates due to the ever-changing thermal environment. Due to the early Sun's enhanced ultraviolet radiation, the photochemical destruction rate during the Wild Years was a factor of 100 higher than the present-day rate, but this only results in a loss of ∼10 m global equivalent layer (GEL). The enhanced Jeans escape rate varies wildly with time, and a net loss of ∼100 cm GEL is predicted. Additionally, we model the impact history during the migration and find that impacts are a net source, not loss, of $N_2$, contributing ∼100 cm GEL. The 100 cm GEL is 0.1% of the amount of $N_2$ in Sputnik Planitia. We therefore conclude that Pluto did not lose an excessive amount of volatiles during the Wild Years, and its primordial volatile inventory can be approximated as its present-day inventory. However, significant fractions of this small total loss of $N_2$ occurred during the Wild Years, so estimates made using present-day rates will be underestimates.

*Unified Astronomy Thesaurus concepts:* Pluto (1267); Plutinos (1266); Planetary migration (2206); Atmospheric evolution (2301); Planetary atmospheres (1244)

## 1. Introduction

Pluto's atmosphere is more than 99% nitrogen (Stern et al. 2015), and much of the observed surface is covered with nitrogen ice, most notably the 1000 km wide ice sheet known as Sputnik Planitia. Observations from the New Horizons mission provide estimates of the present-day nitrogen inventory, which is dominated by the ice in Sputnik Planitia. However, over the course of Pluto's 4.5 Gyr lifetime, various loss mechanisms have been active, and thus Pluto's primordial inventory may have been very different from its present-day inventory.

Modeling the loss of volatiles, specifically $N_2$, from Pluto's surface over the age of the solar system allows estimation of its primordial volatile inventory by calculating the amount lost via escape, impact erosion, and photochemical destruction. These models must account for the formation location and subsequent orbital evolution of Pluto. Pluto is thought to have formed not at 40 au but instead likely closer to the Sun and was pushed outward by interactions with Neptune during giant planet migration, ultimately being captured into the 3:2 mean-motion resonance with Neptune. Models of volatile loss must also account for the variable behavior of the Sun over its lifetime. When the Sun entered the main sequence, its output in ultraviolet (UV) wavelengths, which are important for escape, was orders of magnitude larger than the present-day value. Additionally, the timing of Pluto's differentiation is important for understanding when there were or were not significant deposits of $N_2$ on the surface available to form an atmosphere and be subject to loss mechanisms. In this section, these topics are summarized as they relate to our model of volatile loss from early Pluto.

Pluto likely formed closer to the Sun than its current orbit (Canup et al. 2021, and references therein). Its location in the 3:2 mean-motion resonance with Neptune (meaning Pluto completes two orbits about the Sun for every three that Neptune completes) is evidence for this. As Neptune's orbit expanded, the heliocentric distance corresponding to resonances with its orbit swept out through the solar system as well. Objects originally on orbits unrelated to Neptune could have been "picked up" into the resonance and then continued to evolve in step with Neptune such that they remained in the resonance as it expanded outward from the Sun (Malhotra 1993). The discovery of other objects in Neptune's 3:2 resonance (the so-called "Plutinos") has strengthened this argument (e.g., Yu & Tremaine 1999). Many authors have modeled the giant planet migration. For this work, we use the output from the model of Nesvorný (2015), which constrained the migration of giant planet Neptune specifically based on orbital properties of the Kuiper Belt. From Nesvorný (2015), we have orbital evolution paths for 53 Plutinos, covering a period of 100 Myr in the early part of the solar system. These paths start at the time of the giant planet instability, which probably occurred within the first few tens of millions of years after the protoplanetary disk dispersed. We discuss the exact timing further below.

The Sun is thought to have formed in a giant molecular cloud around 4.6 Gyr ago. The proto-Sun grew in mass as it gained material from the surrounding cloud. Once the proto-Sun stopped gaining mass, it transitioned into what is called a "pre-main-sequence" star. During this stage, the Sun's luminosity was powered by gravitational collapse. For the first part of the stage, known as the T Tauri phase, there was a thick protoplanetary gas disk and strong stellar winds. Sometime later, the Sun condensed enough under its own gravity for hydrogen burning to begin. This point is known as the zero-age main sequence (ZAMS), as it is the beginning of the Sun's time on the main sequence of the H-R diagram. According to the solar evolution model we use here, the pre-main-sequence stage lasted for approximately 40 Myr (Bressan et al. 2012).

When the Sun was only a few tens of Myr old, its radiation output was very different from present-day values; 4.6 Gyr ago, the bolometric luminosity of the "faint young Sun" was just 70% of the present-day value (Bressan et al. 2012). However, at UV wavelengths, the young Sun was actually significantly brighter than it is today. Ribas et al. (2005) measured UV fluxes from six solar analogs of various ages (0.1–7 Gyr) and created a solar evolution model for wavelengths between 1 and 1200 Å, as well as various line fluxes, including Ly$\alpha$ at 1216 Å. Ly$\alpha$ contributed 20% of the total flux between 1 and 1700 Å when the Sun was 0.1 Myr old. Considering the combination of the 1–1200 Å flux plus the Ly$\alpha$ flux, the Sun was 50 times brighter in the UV at 0.1 Gyr old than it is currently at 4.56 Gyr old.

While the early Sun was evolving as described above, the solar system was forming as well. Calcium-aluminum-rich inclusions (CAIs) are small inclusions found inside of chondritic meteorites. They are thought to be among the oldest solid materials in the solar system and are commonly used to determine the solar system's age. Four CAIs from meteorites have been radiometrically dated, yielding an age of $4567.30 \pm 0.16$ Myr (Connelly et al. 2012). CAIs are likely to have condensed around the time the proto-Sun was transitioning into its pre-main-sequence stage, so we assume these two events were concurrent in this work.

The lifetime of protoplanetary disks is needed in order to pin down when the giant planet instability occurred relative to the age of the Sun. In their observational $L$-band study of young to intermediate-age (0.3–30 Myr) stellar clusters, Haisch et al. (2001) found an overall disk lifetime for stars in the clusters they studied of 6 Myr (meaning all of the stars older than 6 Myr did not have disks) and found that half of the stars lost their disks by an age of 3 Myr. The review by Williams & Cieza (2011) establishes an upper limit of 10 Myr for disk lifetimes based on a variety of Spitzer observation programs. Weiss et al. (2021) constrain the lifetime of the Sun's solar nebula, in the outer solar system specifically, to less than 5 Myr based on paleomagnetism arguments. Thus, the Sun's protoplanetary gas disk had likely dissipated fully by 4.557 Ga (10 Myr after CAI condensation), perhaps even earlier. Lisse et al. (2021) discuss the shadowing and therefore cooling effect that the protoplanetary disk had on the Kuiper Belt early in solar system history, but given that the disk disperses before the time period relevant to this work, we do not need to account for this effect.

When originally proposed, a late giant planet instability, occurring ∼700 Myr after the solar system formed, was favored in order to explain the Late Heavy Bombardment suggested by large lunar impact basins (Gomes et al. 2005). However, more recent work argues that the instability occurred much earlier, within 100 Myr of solar system formation. In order to prevent dynamically overexciting the terrestrial planet orbits, the giant planet instability needs to occur within 50 Myr of solar system formation (Nesvorný 2018). For this work, we assume the giant planet instability occurred 30 Myr after solar system formation, defined as the condensation time of CAIs. After the onset of the instability, Neptune's migration over the subsequent ∼100 Myr influenced the orbits of the objects in the proto-Kuiper Belt, eventually capturing some into the 3:2 resonance.

Finally, we need to know when Pluto's $N_2$ inventory reached the surface and became available to loss mechanisms. Differentiation would bring the lighter volatiles to the surface while the heavier rocky material forms a core. Based on the composition and formation models for Charon, it is likely that the Pluto and Charon progenitors were at least partially differentiated at the time of the Charon-forming impact, if not fully differentiated (Canup et al. 2021). Heat from the impact or subsequent tidal heating from the newly formed Charon could finish differentiating Pluto and bring its volatiles, including $N_2$, to the surface (McKinnon et al. 2016). In this work, we assume that either Pluto's nitrogen was originally present as $N_2$ (rather than other chemical species) or the original chemical form of nitrogen was processed into $N_2$ by the start of the Wild Years. This assumption means that the maximum amount of $N_2$ was present and exposed to loss mechanisms at the surface for the maximum amount of time. Tidal heating would be limited to within 1–10 Myr of the Charon-forming impact, after which the orbits would have been circularized (Canup et al. 2021; McKinnon et al. 2021). In order to form Charon from a mostly intact progenitor, as is suggested, the impact velocity needs to be slow, roughly less than 1.2 times the escape velocity. This constraint is best met after the dispersal of the protoplanetary gas disk but before the onset of Neptune migration. Thus, we can conclude that Pluto's $N_2$ inventory was at the surface and subjected to loss mechanisms during the time of Neptune migration.

Figure 1 summarizes the preceding discussion about the relative and absolute timing of events in the early solar system and indicates the assumptions made for this work.

Section 2 goes into more detail about the models of the early Sun, solar system, and Pluto. In Section 3, we present results about the orbital migration itself and the general climates experienced by migrating Plutinos. Sections 4–6 present first our models for each loss mechanism and then the results from volatile loss due to photochemical destruction, impact erosion, and atmospheric escape, respectively. Finally, in Section 7, we discuss what these results imply for Pluto's formation and hypothesized epochs of ancient glaciation.

## 2. Early Solar System, Sun, and Pluto Models

### 2.1. Orbital Migration

As an input to the model, we use the output Plutino orbital evolutions from the dynamical model presented in Nesvorný (2015). In this model, Neptune begins on an orbit with a semimajor axis between 20 and 30 au, and a disk of planetesimals extends from the orbit of Neptune to 30 au. As Neptune migrates, some of the planetesimals are swept up into the 3:2 mean-motion resonance with Neptune, forming the Plutino population. For an ensemble of 53 of these simulated Plutinos, we have the semimajor axis, orbital eccentricity, and inclination at 10,000 yr time steps for a total run time of 100 Myr, covering a period of time we term the "Wild Years."

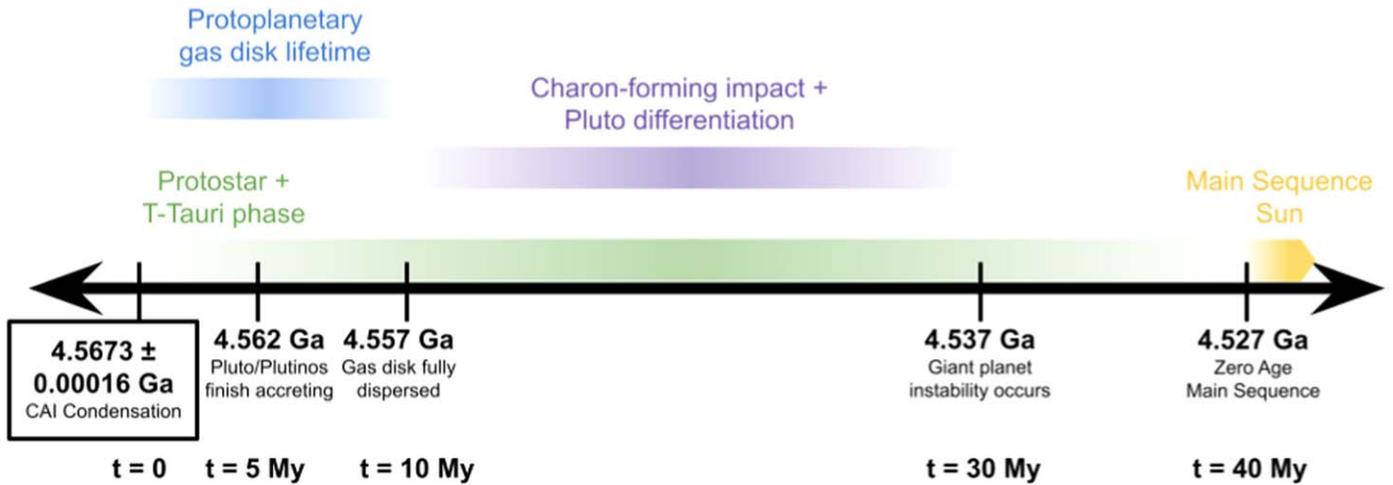

**Figure 1.** Timeline of the early solar system. The CAI condensation time of 4.5673 ± 0.00016 Ga (boxed) is well known via radiometric dating (Connelly et al. 2012). The other times shown on the timeline are more uncertain and represent our assumptions based on the arguments presented in the text. The times shown in the lower row represent time since CAI condensation, which we use throughout the text and figures of this work.

For more details on the creation of these orbits, see Nesvorný (2015). It is important to note that we use these Plutino migrations as a proxy for the possible migration that Pluto specifically underwent, and we use the physical properties of Pluto throughout this work. Figure 2 shows the eccentricity versus semimajor axis migration for all 53 of the Plutinos in our sample. The Plutino labeled 47 is used as an example throughout this paper.

*2.2. Solar Luminosity and UV Flux*

The Sun was less bright overall when it was young. At the time of the Wild Years, the Sun's luminosity was about 70% of the present-day value, and it has steadily increased since then. We use the PARSEC model from Bressan et al. (2012) and the updates from Chen et al. (2014) to calculate the Sun's bolometric luminosity as a function of time, shown in the top panel of Figure 3. The PARSEC model includes the Sun's luminosity behavior during its pre-main-sequence phase, which we assume begins concurrently with CAI condensation and lasts for around 44 Myr. At the ZAMS (44 Myr after CAI condensation), the solar luminosity predicted by the model is $2.7 \times 10^{26}$ W, which is 70% of the present-day value of $3.828 \times 10^{26}$ W.

While the Sun was bolometrically fainter during the Wild Years, the UV flux of the Sun was far stronger during this time period relative to the present day. We use the solar evolution model of Ribas et al. (2005) to estimate the UV flux from the early Sun. The Ribas et al. (2005) model uses observations of seven solar-type stars (including the Sun) of varying ages to create power-law model fits of the stellar flux in various wavelength bands as a function of time.

For part of this work, we are interested in photochemical destruction of $N_2$ molecules in the atmosphere by UV radiation. Radiation at wavelengths between roughly 100 and 1000 Å is capable of breaking the $N_2$ triple bond. We use three model fits from Ribas et al. (2005) in order to cover this wavelength range:

$$F_1(t) = 13.5 t^{-1.2} \quad 100\text{--}360 \text{ Å}, \tag{1}$$

$$F_2(t) = 4.56 t^{-1} \quad 360\text{--}920 \text{ Å}, \tag{2}$$

$$F_3(t) = 2.53 t^{-0.85} \quad 920\text{--}1180 \text{ Å}. \tag{3}$$

The time $t$ is the solar age in Gyr relative to the ZAMS, and the resulting fluxes $F_{1,2,3}$ (at 1 au) are in units of mW m$^{-2}$. We calculate the UV flux enhancement relative to the present-day solar UV flux at a given time with the following:

$$E_{\text{UV}}(t) = \frac{F_1(t) + F_2(t) + F_3(t)}{F_1(t_{\text{present}}) + F_2(t_{\text{present}}) + F_3(t_{\text{present}})}. \tag{4}$$

Additionally, for the energy-limited escape discussed in Section 6.1.1, the early Sun's Ly$\alpha$ flux is also needed. From Ribas et al. (2005), the Sun's Ly$\alpha$ flux as a function of time can be estimated using

$$F_{\text{Ly}\alpha} = 19.2 t^{-0.72}, \tag{5}$$

where once again $t$ is in Gyr and the resulting flux at 1 au is in mW m$^{-2}$. The corresponding enhancement, due to both 100–1180 Å and Ly$\alpha$ radiation, is shown in the middle panel of Figure 3 and given by

$$E_{\text{UV}+\text{Ly}\alpha}$$
$$= \frac{F_1(t) + F_2(t) + F_3(t) + F_{\text{Ly}\alpha}(t)}{F_1(t_{\text{present}}) + F_2(t_{\text{present}}) + F_3(t_{\text{present}}) + F_{\text{Ly}\alpha}(t_{\text{present}})}. \tag{6}$$

*2.3. Radiogenic Internal Heat*

Pluto's radiogenic internal heat flux was also different during the time period of the Wild Years. We implement the model of Hussmann et al. (2010) to calculate radiogenic heat production rates as a function of time $t$:

$$F_{\text{internal}} = \frac{M_{\text{core}}}{4\pi R_{\text{Pluto}}^2} \sum_{i=1}^{4} C_i \mathcal{H}_i e^{-\ln(2) t/\tau_i}, \tag{7}$$

where $M_{\text{core}} = 8.89 \times 10^{21}$ kg is the mass of Pluto's core (assuming a radius of 858 km and a density of 3360 kg m$^{-3}$; Keane et al. 2016), $R_{\text{Pluto}} = 1189$ km is the radius of Pluto, $C_i$ is the abundance of the given isotope based on carbonaceous chondrites, $\mathcal{H}_i$ is the heat release, and $\tau_i$ is the half-life for each isotope. The values we assume for this work are shown in Table 1.

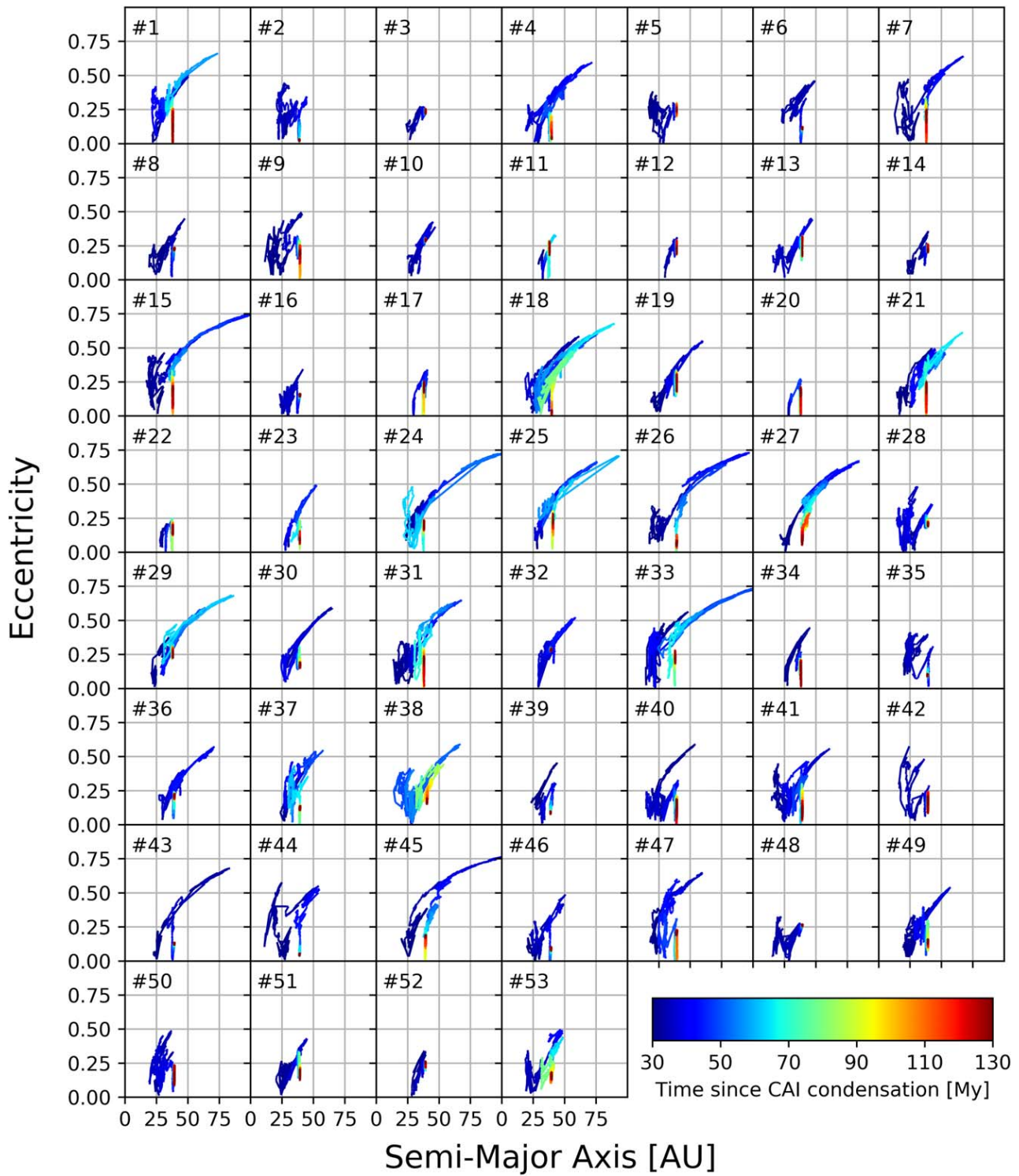

Figure 2. Eccentricity vs. semimajor axis evolution for all 53 of the Plutinos in the sample. The color scale indicates time since CAI condensation.

In Equation (7), we divide by the total surface area of Pluto and multiply by the mass of Pluto's core because we assume that all of the heat produced from radioactive decay in the core exits the body as a surface heat flux and none of the heat remains in the body to change its temperature. In this way, the surface heat flux we use is an upper limit to the true heat flux that could result from this radioactive decay. The bottom panel of Figure 3 shows the resulting surface heat flux as a function of time, which varies from 15 mW m$^{-2}$ at the start of the Wild Years to 12 mW m$^{-2}$ at the end. The present-day surface heat flux is estimated to be a few mW m$^{-2}$ (McKinnon et al. 2016).

### 2.4. Climate Model

To calculate the surface temperature as a function of time, we use the VT3D model (Young 2012, 2017). Here we

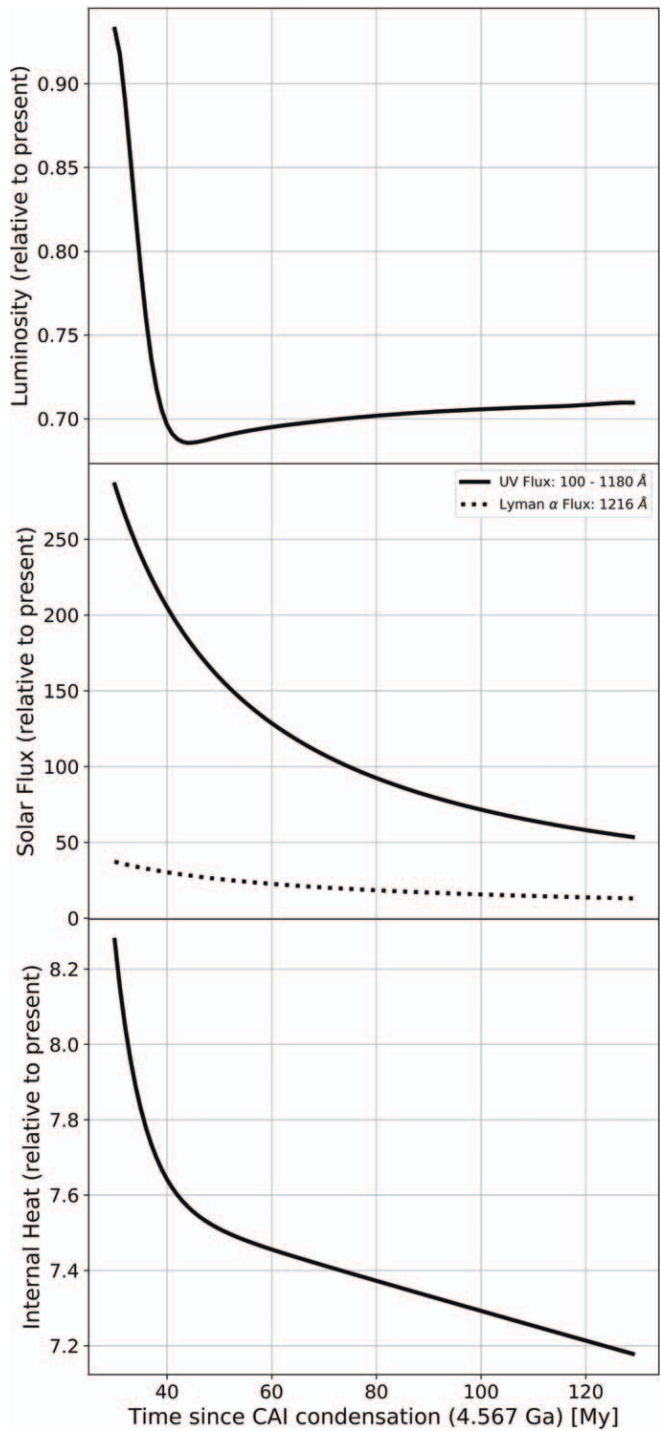

**Figure 3.** (Top) Bolometric luminosity of the Sun during the orbital migration period relative to the present-day luminosity. (Middle) UV and Ly$\alpha$ flux of the Sun during the orbital migration period relative to the present-day UV and Ly$\alpha$ fluxes. (Bottom) Radiogenic internal heat flux of Pluto during the orbital migration period relative to the present-day rate.

**Table 1**
Values for the Abundance $C$, Heat Release $\mathcal{H}$, and Half-life $\tau$ of Each of the Seven Isotopes Used to Calculate Radiogenic Internal Heating

| Isotope | | $C$ | $\mathcal{H}$ | $\tau$ |
| --- | --- | --- | --- | --- |
| | | (ppb) | (W kg$^{-1}$) | (Myr) |
| Long-lived isotopes | $^{238}$U | 19.9 | $94.65 \times 10^{-6}$ | 4468 |
| | $^{235}$U | 5.4 | $568.7 \times 10^{-6}$ | 703.81 |
| | $^{232}$Th | 38.7 | $26.38 \times 10^{-6}$ | 14,030 |
| | $^{40}$K | 738 | $29.17 \times 10^{-6}$ | 1277 |
| Short-lived isotopes | $^{26}$Al | 600 | 0.146 | 0.73 |
| | $^{60}$Fe | 225 | 0.074 | 1.5 |
| | $^{53}$Mn | 25.7 | 0.027 | 3.7 |

**Note.** The long-lived isotope values are from Robuchon & Nimmo (2011), and the short-lived isotope values are from Castillo-Rogez et al. (2007).

summarize the relevant parts of VT3D to this work. VT3D is an energy balance model, finding the temperature that results from the balance between thermal emission, solar insolation, internal heat, thermal conduction, and latent heat of sublimation. For this work, we assume that Pluto is in a "snowball" state: uniformly covered by N$_2$ ice for the duration of the Wild Years. The timing is highly uncertain for the impact that created Sputnik Planitia, and by assuming uniform N$_2$ ice coverage, we implicitly assume it occurred after the Wild Years ended. We assume the ice has an albedo of 0.8, a thermal inertia of 1225 "thermal inertia units" (tiu; equivalent to J m$^{-2}$ K$^{-1}$ s$^{1/2}$), and an emissivity of 0.6 (these values are similar to Reference Case A from Johnson et al. 2021b). Sputnik Planitia has a Bond albedo of 0.8–1 (Buratti et al. 2017), which we took to be a typical albedo for thick N$_2$ ice deposits. The orbital migrations from Nesvorný (2015) are spaced every 10,000 yr. We use the analytic form of VT3D to calculate the nitrogen ice temperature at 20 equally spaced time steps in each of these distinct orbits. We found that 20 time steps was sufficient to resolve the extrema of the orbit while minimizing computation time. Using the current orbit's semimajor axis and eccentricity, we calculate the absorbed solar insolation at each time step, subject to the "faint young Sun" behavior discussed above, as well as the internal heat flux at the given time. This annual temperature behavior is repeated for as many orbits as can be completed in the 10,000 yr macro time step, given the current orbital period.

### 3. Migration and Climate Results

#### 3.1. Orbital Migration during the Wild Years

Figure 4 shows a typical Plutino's path throughout the solar system during the Wild Years, both in eccentricity versus semimajor axis space (left panel) and in semimajor axis, perihelion, and aphelion distances as a function of time (right). This Plutino is shown as #47 in Figure 2. Typical Plutinos reach minimum heliocentric distances of 15 au and maximum heliocentric distances of 80 au, although distances greater than 150 au are possible for certain Plutinos for short periods of time (<50,000 yr). This period of rapid orbital evolution is short-lived, typically lasting only 30 Myr or so, before the Plutino settles down into an orbit near Pluto's present-day orbit. Figure 5 shows a whisker plot of the time spent in each heliocentric distance bin, showing that the bulk of the time (80 Myr out of the total 100 Myr simulation) is spent in heliocentric distances between 30 and 50 au (Pluto's present-day heliocentric distance range). For the remainder of the simulation time that the average Plutino spends outside of the 30–50 au range, most of it is spent closer to the Sun. The average Plutino spends 9 Myr at heliocentric distances less than 30 au and 1.5 Myr at heliocentric distances larger than 50 au.

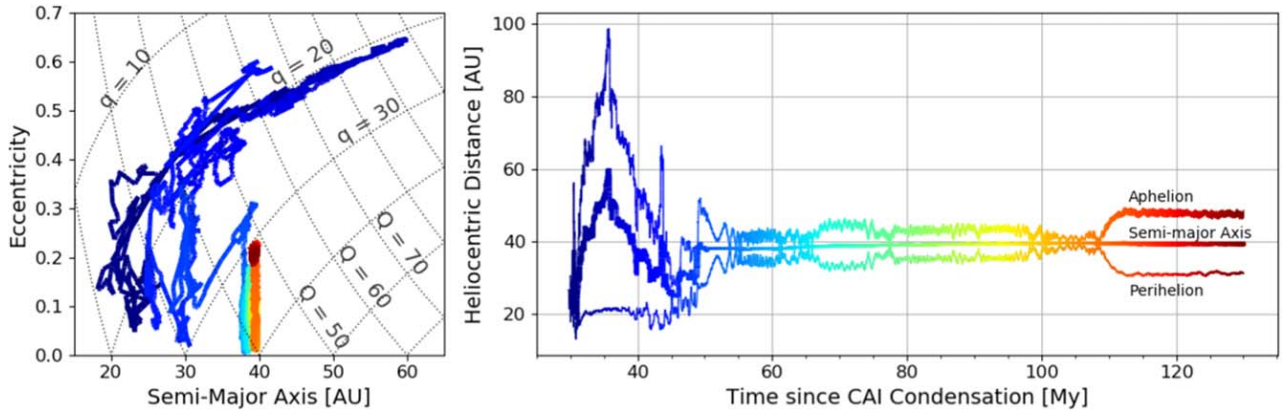

**Figure 4.** Orbital migration for an example Plutino. (Left) Eccentricity vs. semimajor axis, with color indicating the time since simulation start (see the right panel for color scale). Several lines of constant perihelion $q$ and aphelion $Q$ are labeled. (Right) Aphelion and perihelion distances vs. time since simulation start.

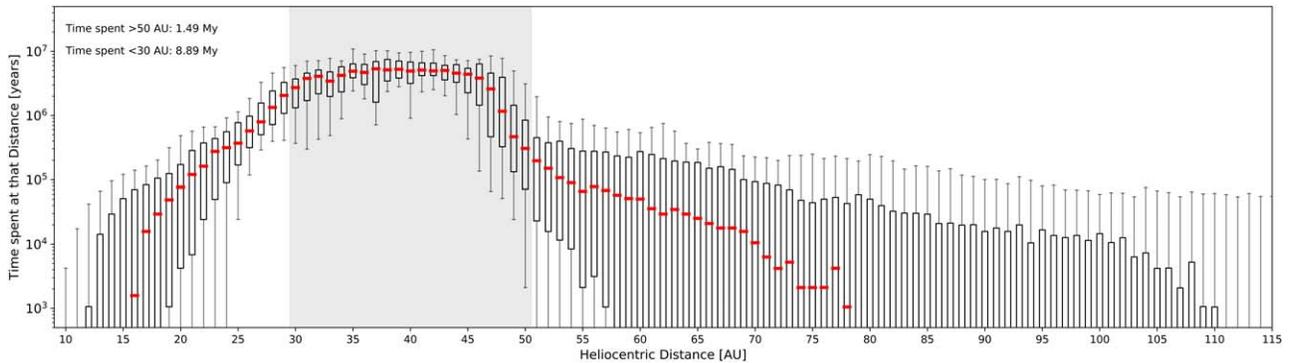

**Figure 5.** Whisker plot showing the time spent at a given heliocentric distance for the ensemble of Plutinos. The red lines are the median value, the boxes encompass the 25th–75th percentiles, and the whiskers are the 5th and 95th percentiles. For the average Plutino, 80 Myr out of the 100 Myr simulation is spent at heliocentric distances between 30 and 50 au (shaded region), which is Pluto's present-day orbital range.

### 3.2. Climate during the Wild Years

Before discussing volatile loss rates, we start by showing the surface temperature and surface atmospheric pressure experienced by the Plutinos during their orbital migration to give an idea of the range in general climate characteristics. Figure 6(B) shows the temperature evolution of an example Plutino (the same Plutino from Figure 4). The surface temperature of the Plutino is calculated using the VT3D model (as discussed in Section 2.4) with uniform spatial coverage of nitrogen ice with an albedo of 0.8, a thermal inertia of 1225 tiu, and an emissivity of 0.6. This particular Plutino begins in an orbit with a semimajor axis of 24 au, but its orbit quickly expands to a semimajor axis near 60 au and an eccentricity as high as 0.65, leading to aphelion values near 100 au. As this orbit expansion occurs, the surface temperature falls from initial values as high as 53 K to minimum values of 32 K. Subsequently, this Plutino's orbit slowly recontracts to a final semimajor axis of 39.5 au (after overshooting slightly), and the surface warms accordingly to a final surface temperature of 35–36 K. This final surface temperature is roughly 1 K cooler than the present-day surface temperature of a Pluto-like object in this orbit with these thermal parameters would be, due primarily to the faint young Sun luminosity (see Figure 3). The internal heat is higher during this time period relative to the present day, but the solar luminosity has a larger effect on temperature because the absorbed insolation at the present day is 57 mW m$^{-2}$ versus the present-day internal heat flux of only 2 mW m$^{-2}$.

Figure 7 shows histograms of the time spent at each temperature for the suite of Plutinos. The red lines indicate the median time spent at that temperature, the boxes encompass the 25th–75th percentiles, and the whiskers indicate the 5th and 95th percentiles. The shaded region (35–36 K) highlights the temperature range of the Plutinos once they reach their final orbit near 40 au; all the Plutinos spend the majority of the simulation time (tens of Myr) in this temperature range. For context, models suggest that Pluto's mean surface temperature might seasonally vary between 33 and 37 K in its present-day orbit, depending on assumptions about its present-day nitrogen distribution (Johnson et al. 2021b). If Pluto is assumed to be uniformly covered in N$_2$ ice in order to match the assumption made here, its surface temperature would vary between 36 and 37 K over its present-day orbit.

The lowest temperature reached is ~28 K, typically for less than 10,000 yr. The surface pressure in equilibrium with nitrogen ice at 28 K is 4.4 nbar (Fray & Schmitt 2009). Johnson et al. (2015) calculated that for a nitrogen atmosphere to have an optical depth of 1 in UV wavelengths, a minimum column abundance $N_C = 2 \times 10^{18}$ cm$^{-2}$ is required. This column abundance implies a surface pressure of $N_C \times m_{N2} \times g_{surf} = 5.8$ nbar (implying a surface temperature of 28.2 K, assuming vapor pressure equilibrium), shown as the blue dotted line in Figure 7. So at the very lowest temperatures, it is possible for UV rays to reach all the way to the surface of the Plutinos, albeit only for 10,000 or fewer years cumulative during the Wild Years period. This calculation of a UV-transparency limit assumes that the $CH_4$ mixing ratio in Pluto's atmosphere is constant throughout time. If the $CH_4$ mixing ratio is lower at lower temperatures (Tan 2022), then the integrated duration could be higher, but likely still small.

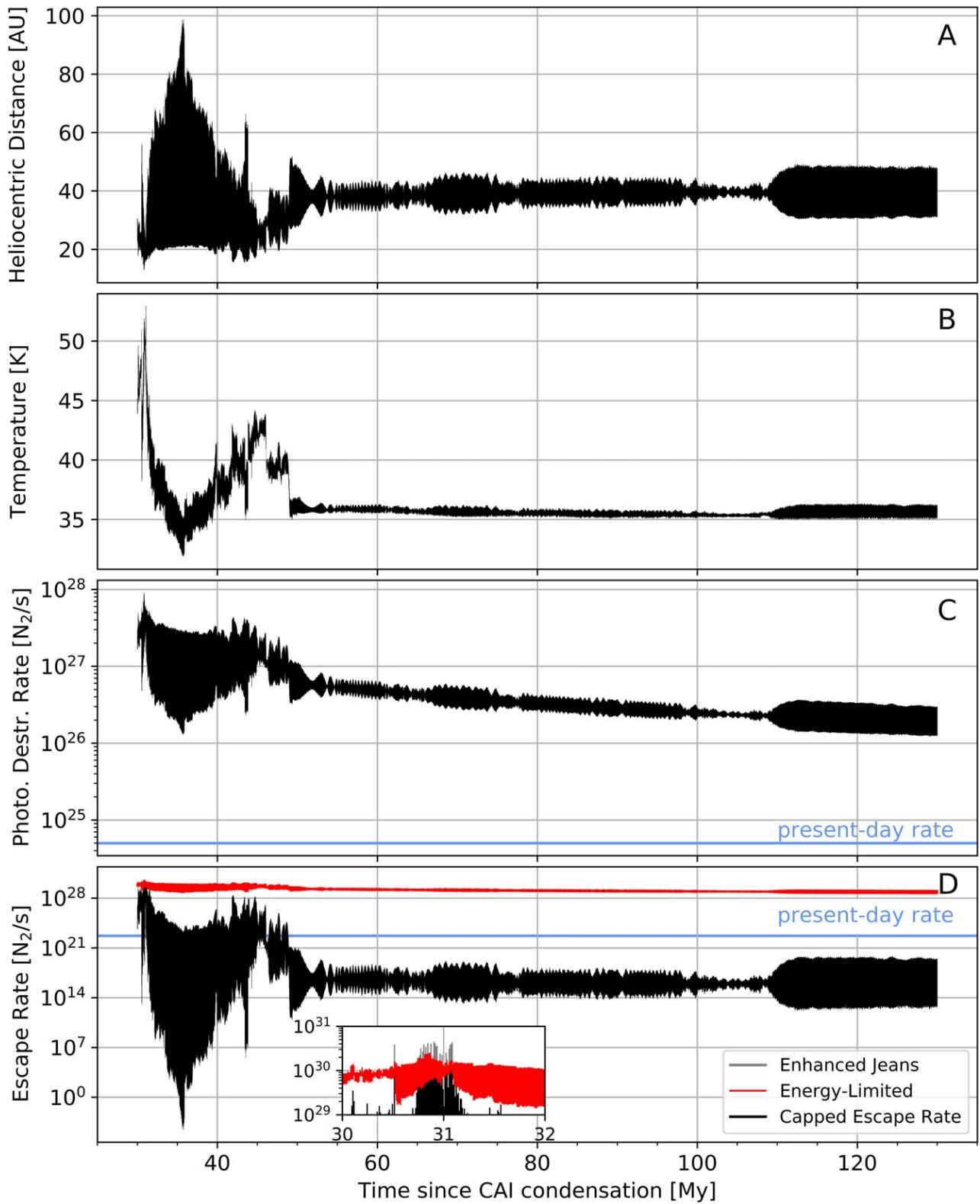

**Figure 6.** Panel (A): heliocentric distance as a function of time for the example Plutino during its orbital migration. See Figure 4 for the orbital parameters of this Plutino as a function of time. Panel (B): temperature vs. time for an example Plutino during its orbital migration. By ∼50 Myr after CAI condensation, the Plutino has been captured into resonance with Neptune and the subsequent orbit changes are minimal, so the surface temperature is nearly constant in time. Seasonal temperature changes are on the order of 1 K and contribute to the apparent thickness of the line. Panel (C): photochemical destruction rate as a function of time for the example Plutino. The present-day photochemical destruction rate from Krasnopolsky (2020) is shown as the horizontal blue line. We attribute the general decreasing trend to the decrease in solar UV output over time, which falls from 250 times the present-day value to only 50 times the present-day value over this time period (Ribas et al. 2005). Panel (D): escape rate as a function of time for the example Plutino. The red line shows the energy-limited escape rate. The enhanced Jeans escape rate (gray) exceeds the energy-limited value near the start of the Wild Years, and only barely (see inset for zoomed-in view). The present-day escape rate (Strobel 2021) is shown in blue.

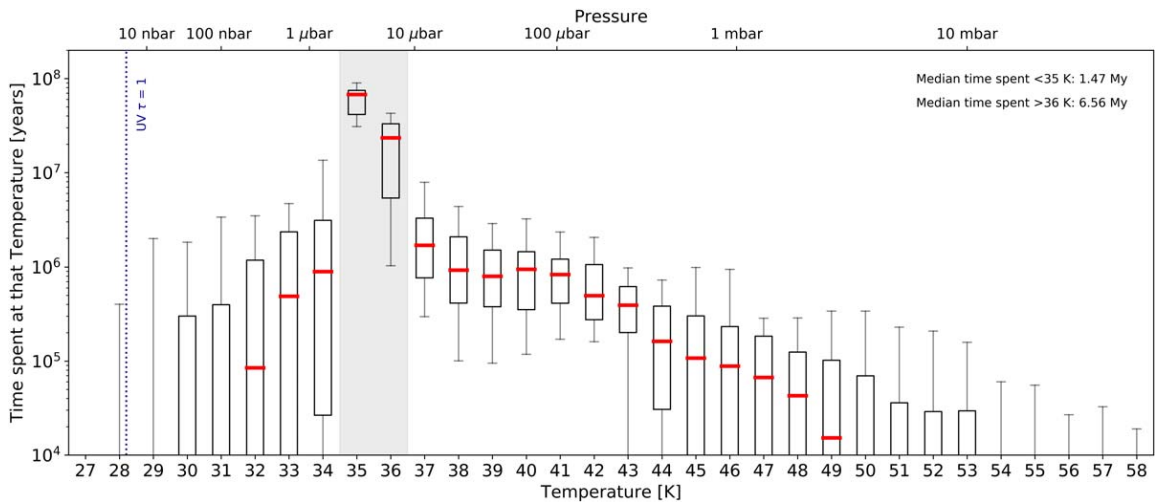

**Figure 7.** Histograms of the time spent at a given temperature during the 100 Myr Wild Years period. The *x*-axis spans the full range of temperatures experienced by the ensemble of Plutinos during the Wild Years. The red lines are the median value, the boxes encompass the 25th–75th percentiles, and the whiskers are the 5th and 95th percentiles. For temperatures with no red median line shown in the figure, the median Plutino spends zero time at that temperature. The shaded region (35–36 K) indicates the temperatures that Plutinos experience once they reach their final orbits near 40 au. The top axis shows the vapor pressure that is in equilibrium with nitrogen ice at the given temperature. All the Plutinos spend the bulk of their time (tens of Myr) at 35–36 K.

The highest temperature reached by the median Plutino is around 48 K, corresponding to a Mars-like surface pressure of 2.1 mbar. The Plutinos are at this temperature for less than $10^5$ yr, typically. The triple point for $N_2$ is 63 K (Fray & Schmitt 2009), so liquid $N_2$ is never stable on Pluto's surface even at the hottest points during the Wild Years. Stern et al. (2017) proposed that the observed dendritic valleys on Pluto could have been carved by liquids and identified Alconis Lacus as a potential paleo-lake, but our results show that temperatures likely do not get high enough to support these hypotheses. We discuss the alternative possibility that glaciers could have carved the valleys in Section 7.

## 4. Photochemical Destruction

One loss mechanism for nitrogen from Pluto's atmosphere is the irreversible photochemical destruction of $N_2$ in the atmosphere, which primarily creates solid-phase daughter products that fall out of the atmosphere onto the surface, although some escape to space is possible as well. Krasnopolsky (2020) used a photochemical model to calculate the loss of atmospheric $N_2$ due to photochemical destruction at the time of the New Horizons flyby of 37 g cm$^{-2}$ Gyr$^{-1}$, equivalent to $5 \times 10^{24}$ $N_2$ s$^{-1}$, which is roughly 100 times greater than the atmospheric escape rate at the time of the flyby (Strobel 2021). The dominant daughter product is HCN.

In present-day Pluto, the majority of the photochemical loss of $N_2$ is in the form of nitriles such as HCN, so the loss rate is dependent on having sufficient $CH_4$ in the atmosphere as well to provide CH and $CH_3$ radicals (Krasnopolsky 2020). As observed by New Horizons, there is copious $CH_4$ available in Pluto's upper atmosphere (roughly 1% at the altitudes of $N_2$ dissociation; Young et al. 2018). Unlike Triton during the Voyager 2 flyby, where photochemical destruction removed $CH_4$ at high altitudes, the $CH_4$ mixing ratio is large in Pluto's current atmosphere because diffusive separation enhances the lighter $CH_4$ at high altitudes. Pluto's $CH_4$ mixing ratio at the surface is controlled by vapor pressure equilibrium with the surface ices in a manner not yet completely understood (Young et al. 2021), which may be related to warm areas of methane-rich ices. On an iceball Pluto, as we model here, the $CH_4$ mixing ratio at the surface may well be lower than that of present-day Pluto. Regardless, since Pluto has so much more $CH_4$ than Triton, in this work, we make the assumption that vapor pressure equilibrium and diffusive separation lead to plentiful $CH_4$ at the altitudes of $N_2$ dissociation throughout the Wild Years and all of Pluto's history. This assumption could be revisited once the problem of Pluto's high surface $CH_4$ mixing ratio has been explained for the present-day atmosphere. If instead there is not always sufficient $CH_4$ in Pluto's atmosphere, the $N_2$ photochemical destruction rates presented here will likely be upper limits. If the lack of $CH_4$ is due to a complete removal of $CH_4$ ice from the surface, then it is possible that the atmosphere may cool significantly and collapse, reducing the $N_2$ photodestruction rate. An alternate option for a lack of atmospheric $CH_4$ is due to the Hunten limiting flux preventing sufficient $CH_4$ from reaching the upper atmosphere altitudes (Hunten 1973). In this case, $CH_4$ could still be photolyzed near the surface, and the $N_2$ photodestruction rate may drop, but likely not as much as the first case. We do not treat either of these scenarios carefully in the following analysis and instead make the assumption that there is always sufficient $CH_4$ for $N_2$ photodestruction to occur, such that the photodestruction rate is dependent on the incident solar radiation alone.

### 4.1. Photochemical Destruction Model

In order to estimate the photochemical destruction rate during the Wild Years, we scale the present-day rate from Krasnopolsky (2020) based on both solar output and heliocentric distance. The "present-day" rate is based on the New Horizons flyby conditions, which occurred when Pluto was at a heliocentric distance of 32.9 au. The photochemical destruction rate $F_{photochem}$ should depend on the inverse square

Table 2
Summary of the Nitrogen Loss, in cm GEL, Resulting from Each of the Loss Mechanisms Investigated Here

| Loss Mechanism | | During Wild Years (30–130 Myr after CAIs) | Age of Solar System (130 Myr after CAIs–present) | Total |
|---|---|---|---|---|
| Photochemical destruction[a] | Baseline size: | $385^{+13}_{-23}$ cm GEL | 885 cm GEL | $1270^{+13}_{-23}$ cm GEL |
| | Variable size: | $44^{+47}_{-9}$ cm GEL | 180 cm GEL | $224^{+47}_{-9}$ cm GEL |
| Impact delivery[b] | | $-(74^{+45}_{-8})$ cm GEL | $-(74^{+45}_{-8})$ cm GEL | $-(168^{+83}_{-28})$ cm GEL |
| Impact removal[b] | Hot atmo.: | $100^{+19}_{-37}$ cm GEL | $10^{+5}_{-4}$ cm GEL | $110^{+25}_{-37}$ cm GEL |
| | Present−day atmo.: | $10^{+5}_{-4}$ cm GEL | | $22^{+6}_{-6}$ cm GEL |
| | Cold atmo.: | $1.8^{+0.7}_{-0.8}$ cm GEL | | $12.5^{+4.7}_{-4.1}$ cm GEL |
| Jeans escape[a] | | $98^{+773}_{-97}$ cm GEL | 2.35 cm GEL | $100.35^{+773}_{-97}$ cm GEL |

**Notes.**
[a] The subscript and superscript values represent the 25th and 75th percentile ranges based on the ensemble of 53 Plutinos for the photochemical destruction and Jeans escape rows. The values for the age of the solar system column do not have reported uncertainties because we use Pluto's present-day orbit for that time period.
[b] The subscript and superscript values represent the 25th and 75th percentile ranges based on the 100 random samples drawn for the impact delivery and removal rows. The apparent discrepancy between the sum of the "During Wild Years" and "Age of Solar System" columns and the value reported in the "Total" column is due to the random draw of impactors. The medians and 25th and 75th percentiles cannot be simply added as they were for the photochemical and escape rows.

of heliocentric distance $h$; thus, we can write the relation

$$F_{\text{photochem}}(t, h) \propto \left(\frac{32.9 \text{ au}}{h}\right)^2. \quad (8)$$

Additionally, the photochemical destruction rate will depend on the solar output at the given time. As discussed above, the solar output was very different during the Wild Years as compared to the present-day output. $N_2$ is photochemically destroyed by radiation at UV wavelengths, so we scale the present-day rate based on the Ribas et al. (2005) model of early stellar output. For the dependence on solar output, we can write the following relation:

$$F_{\text{photochem}}(t, h) \propto E_{\text{UV}}(t), \quad (9)$$

where $E_{\text{UV}}(t)$ is the UV enhancement at a given time (relative to the present-day solar output) and is calculated using Equation (4) from Section 2. Ly$\alpha$ is not energetic enough to break the bond in a molecule of $N_2$, so we do not account for the early Sun enhancement in Ly$\alpha$ for the photochemical destruction.

By combining the two preceding relationships, we can write an equation for the photochemical destruction rate of $N_2$ in Pluto's atmosphere as a function of both time and heliocentric distance during the Wild Years:

$$\begin{aligned}&F_{\text{photochem}}(t, h)\\&= E_{\text{UV}}(t)\left(\frac{32.9 \text{ au}}{h}\right)^2 F_{\text{photochem}}(t_{\text{present}}, 32.9 \text{ au}),\end{aligned} \quad (10)$$

where $F_{\text{photochem}}(t_{\text{present}}, 32.9 \text{ au}) = 5 \times 10^{24} \text{ N}_2 \text{ s}^{-1}$ is the photochemical destruction rate calculated for Pluto's conditions at the time of the New Horizons flyby in 2015 (Krasnopolsky 2020).

### 4.2. Photochemical Destruction Results

Panel (C) in Figure 6 shows the calculated photochemical destruction rate as a function of time for the example Plutino. For reference, the present-day photochemical destruction rate (Krasnopolsky 2020) is shown as the blue line; note that the calculated photochemical destruction rate during the Wild Years is always at least an order of magnitude higher than the present-day rate, primarily due to the enhanced early solar UV output. The photochemical destruction rate generally decreases with time, following the decreasing trend of the solar UV flux. The solar UV output falls from over 250 times the present-day value to just 50 times the present-day value during the Wild Years. The photochemical destruction rate is near $3 \times 10^{27} \text{ N}_2 \text{ s}^{-1}$ at the start of the simulation time and falls by an order of magnitude to $2 \times 10^{26} \text{ N}_2 \text{ s}^{-1}$ by the end of the simulation due to the decreasing solar UV output and the orbital migration from an orbit with a semimajor axis of 24 au to one with a semimajor axis of 39 au.

The total photochemical destruction of $N_2$ during the Wild Years for the example Plutino amounts to 387 cm global equivalent layer (GEL), and the remainder of the age of the solar system adds another 857 cm GEL. The median values for the entire ensemble are shown in Table 2, which are very similar to this example Plutino's amounts. The Wild Years are only 100 Myr long, accounting for only 2% of the age of the solar system, but we calculate that roughly 30% of the total photochemical destruction of $N_2$ occurred during the Wild Years. This can primarily be attributed to the highly enhanced solar UV output early in solar system history. Pluto's present-day atmosphere, if condensed onto the surface, amounts to 0.2 cm GEL, and the ice sheet inside of Sputnik Planitia basin is estimated to contain 44–440 m GEL of nitrogen ice (McKinnon et al. 2016; Trowbridge et al. 2016; Johnson et al. 2021a). The photochemical destruction amounts we calculate here fall in between these two values, equivalent to the destruction of more than 6000 present-day Pluto atmospheres worth of $N_2$, which is 3%–30% of the amount of nitrogen contained within Sputnik Planitia.

### 4.3. Photochemical Destruction with a Variable Atmosphere Size

In the above treatment of photochemical destruction of $N_2$ in Pluto's atmosphere, we neglected the variable radial extent of the atmosphere throughout the orbit evolution. The radial extent determines how much of the incoming solar flux is absorbed by Pluto's atmosphere by changing the cross-sectional size of the "target." We present one method for parameterizing the radial extent of Pluto's ancient atmosphere, but the large number of unknowns prevents us from making definitive conclusions as to the effect of a variable atmosphere's size on the loss of $N_2$ through this mechanism. We instead draw our conclusions from the photochemical destruction methodology presented in Section 4.1, in which the radial extent of the atmosphere is held constant.

The radial extent of Pluto's atmosphere controls the effective cross section, which in turn determines how much incoming solar radiation is absorbed by the atmosphere. The absorbed radiation is proportional to $\pi r_{\max}^2$. We use the radius where the line-of-sight optical depth of $N_2$ $\tau_{\text{LOS}} = 1$ for the maximum radial extent $r_{\max}$ of the atmosphere, although arguments could be made for the use of different radii. To calculate the line-of-sight optical depth, we use a brute-force approach as follows. At every radius above the surface, we sum up the $N_2$ density (using Equation (24), assuming the atmosphere is isothermal with an altitude-dependent gravitational acceleration) in each grid box along the line of sight to yield a line-of-sight column density, $N_{\text{LOS}}$. This is converted to an optical depth via

$$\tau_{\text{LOS}}(r) = N_{\text{LOS}}(r) \sigma_{\text{photo}}, \quad (11)$$

where $\sigma_{\text{photo}}$ is the photochemical absorption cross section for $N_2$, $1 \times 10^{-22}\,\text{m}^2$ from the PHIDRATES database (Huebner & Mukherjee 2015). The radius $r$ at which $\tau_{\text{LOS}}(r) = 1$ is then taken to be the maximum radial extent $r_{\max}$ of the atmosphere for that particular point in time.

For some combinations of atmospheric parameters experienced by the ensemble of Plutinos, the line-of-sight optical depth never equals 1 at any altitude due to our assumption of an isothermal atmosphere. Essentially, in a hypothetical atmosphere that is isothermal from the surface to infinity, the density does not fall off quickly enough, and the line-of-sight optical depth minimizes at a value larger than 1. For these cases, we adjust the radial density profile to one that decreases faster with altitude at a certain height. Low in the atmosphere, the density is calculated as normal using Equation (24), in which the atmospheric temperature is held constant with radius, but gravity is allowed to vary. This is equivalent to using a density scale height of

$$H_{\text{iso}}(r < r_{2-\text{level}}) = \frac{k_B T}{m_{N_2} g(r)}. \quad (12)$$

At a given radius $r_{2\text{-level}}$, we switch to an atmospheric structure in which density drops off as $r^{-2}$, which yields a scale height of

$$H_{\text{rsquared}}(r > r_{2-\text{level}}) = \frac{r}{2}. \quad (13)$$

The radius at which these two scale heights are equivalent, $r_{2\text{-level}}$, is taken to be the transition between the lower atmosphere with the isothermal structure and the upper atmosphere with pressure dropping off as $r^{-2}$. Using this two-level structure, we calculate the atmospheric density profile, the resulting optical depth profile using Equation (11), and, finally, the radius at which the optical depth equals 1, $r_{\max}$.

Once we calculate the radius $r_{\max}$, we can use this to scale the photochemical destruction rate with atmospheric size. We adjust Equation (10) to add this dependence on $r_{\max}^2$:

$$F_{\text{photochem}}(t, h) = \left(\frac{r_{\max}(t)}{r_{\max}(t_{\text{present}})}\right)^2 E_{\text{UV}}(t)$$
$$\left(\frac{32.9\,\text{au}}{h}\right)^2 F_{\text{photochem}}(t_{\text{present}}, 32.9\,\text{au}). \quad (14)$$

We calculate that $r_{\max}(t_{\text{present}})$, the present-day radial extent of Pluto's atmosphere, is 1714 km.

Using the adjusted photochemical destruction rate equation, we find that the net loss of $N_2$ during the Wild Years is reduced by an order of magnitude. The dependence on atmospheric size means that the photochemical destruction rate is higher than we previously calculated when a Plutino has a hot and extended atmosphere. However, for the bulk of the Wild Years, the radial extent of most Plutinos is actually less than Pluto's present-day atmospheric size, so the photochemical destruction rate is lower than we previously calculated. The net effect is a reduction in the photochemical loss of $N_2$ from an average of 385 cm GEL to 44 cm GEL. We also find that the range of possible loss amounts varies significantly more between Plutinos when we include the variable atmospheric size, compared to the previous results. The 25th and 75th percentile range increased from 36 cm GEL to 56 cm GEL (see Table 2 for details).

## 5. Impact Delivery and Erosion

The next loss mechanism for nitrogen that we investigate here is impact delivery and removal. Impacts can deliver volatiles to Pluto's surface (or Plutinos' surfaces) if the impactors contain volatiles and those volatiles are retained after the impact, either as fragments on the surface or as gas in the atmosphere. However, impacts can also remove volatiles by injecting energy into the atmosphere, which allows particles to escape. The size and frequency of the impactors determines which of these processes dominates and if impactors are a net source or a net loss of volatiles from a body.

Planetary surfaces record their impact history in the form of craters. There are complications, including that new craters can "erase" or overprint old craters on heavily cratered surfaces; geologic activity, such as volcanism, can obscure existing craters by resurfacing; and impacts can occur before a planetary surface has solidified and become capable of recording craters. Morbidelli et al. (2021) combine crater-counting techniques and a dynamical model of the solar system in order to estimate the total number of impacts onto Pluto throughout its history, accounting for these complicating factors. They define one "integrated current impact rate" (ICR) to be the total number of impactors calculated assuming the present-day impact rate applied for the past 4.5 Gyr. However, the impact rate was higher in the past, such that Pluto actually experienced four ICRs worth of impactors in the past 4.5 Gyr according to Morbidelli et al. (2021). Assuming a size–frequency distribution of $N(>d) \propto d^{-2.1}$ (as in Morbidelli et al. 2021), where $d$ is the impactor diameter, and an impactor density of $1000\,\text{kg m}^{-3}$, four ICRs corresponds to $1.16 \times 10^{18}\,\text{kg}$ of material delivered to Pluto. There is some evidence that large KBOs (>50 km) may follow a steeper size–frequency distribution (e.g., Fraser et al. 2014). However, as seen in the bottom panel of Figure 8, impacts that large are already rare in our random sample; a steeper slope would reduce the likelihood of one occurring even further. Additionally, as seen in the top panel, impacts of that size contribute a net delivery of volatiles but remove the same amount as a smaller impactor, since all impactors with $r_{\rm cap} < r < r_{\rm gi}$ remove the same amount of atmosphere (see Equations (16) and (17) for definitions of $r_{\rm cap}$ and $r_{\rm gi}$). Thus, the exact size–frequency distribution slope used for large KBOs does not have a significant impact on the results discussed here.

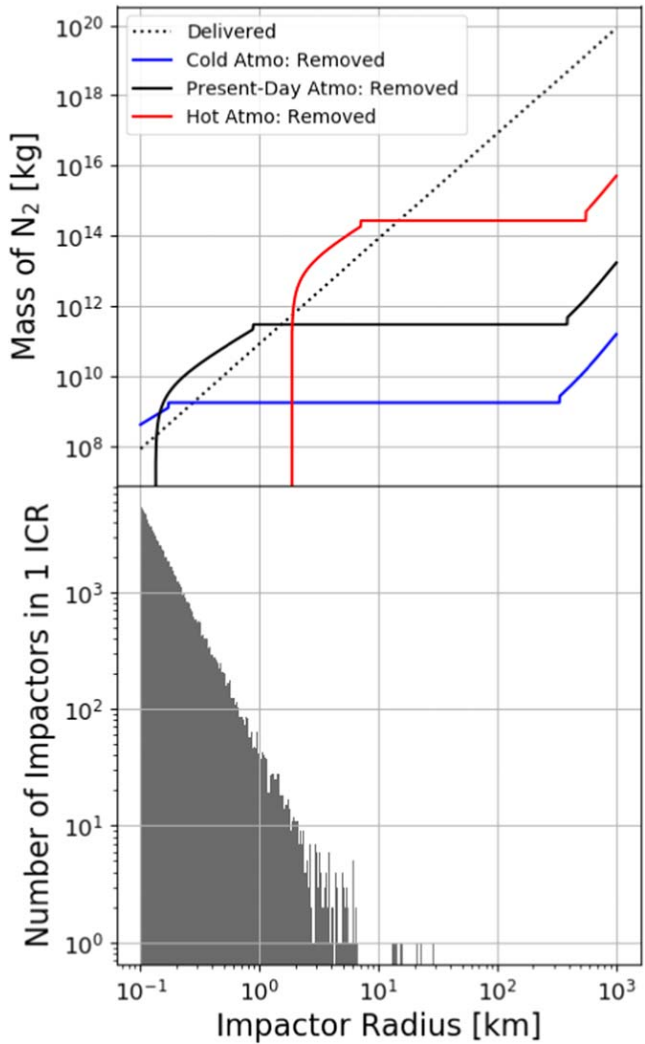

**Figure 8.** (Top) Mass delivered (dotted line) or removed (solid lines) by a single impactor as a function of impactor size. The red line corresponds to the amount removed from a hot atmosphere ($T_u = 124$ K, $H_{\rm surf} = 60$ km, $\rho_{\rm surf} = 10^{-2}$ kg m$^{-3}$), black to Pluto's present-day atmosphere ($T_u = 65$ K, $H_{\rm surf} = 20$ km, $\rho_{\rm surf} = 10^{-4}$ kg m$^{-3}$), and blue to a cold atmosphere ($T_u = 27$ K, $H_{\rm surf} = 13$ km, $\rho_{\rm surf} = 10^{-6}$ kg m$^{-3}$). (Bottom) Histogram showing the number of impactors of a given size in a random sample equivalent to one ICR.

We assume that all of the impactor mass is retained by Pluto after the impact as an upper limit estimate of the amount of N$_2$ delivered by impacts. Considering only nitrogen, and assuming that each impactor is 2% nitrogen in all forms by mass (Mumma & Charnley 2011; Singer & Stern 2015), the total delivery of nitrogen is $2.32 \times 10^{16}$ kg, equivalent to 1000 times the mass of Pluto's present-day atmosphere or 1% of the mass of nitrogen thought to be contained in Sputnik Planitia. Our assumption that impactors are 2% nitrogen by mass may be an overestimate; in that case, our estimation of the amount of nitrogen delivered by impactors is also an overestimate. However, this assumption does not affect the amount of atmospheric nitrogen removed during the impact, which we calculate below.

There are many models of how impacts remove gas from an atmosphere. In this work, we follow the framework of Schlichting et al. (2015) but adapt their models from Earth to Pluto parameters. Schlichting et al. (2015) define four impactor size regimes.

1. $r < r_{\rm min}$: impactors that are too small to remove any atmosphere.
2. $r_{\rm min} \leqslant r < r_{\rm cap}$: impactors that can remove some of the atmosphere local to the impact site.
3. $r_{\rm cap} \leqslant r < r_{\rm gi}$: impactors that remove all of the atmosphere local to the impact site, meaning all of the atmosphere above the tangent plane to the impact site.
4. $r \geqslant r_{\rm gi}$: giant impactors capable of removing all of the local atmosphere and some (or all) of the nonlocal atmosphere material through ground shocks.

The equations for the impactor radii $r_{\rm min}$, $r_{\rm cap}$, and $r_{\rm gi}$ are given below:

$$r_{\rm min} = \left(\frac{3\rho_{\rm atmo}}{\rho_{\rm imp}}\right)^{1/3} H, \quad (15)$$

$$r_{\rm cap} = \left(\frac{3\sqrt{2\pi}\,\rho_{\rm atmo}}{4\rho_{\rm imp}}\right)^{1/3} \sqrt{HR_{\rm Pluto}}, \quad (16)$$

$$r_{\rm gi} = (2HR_{\rm Pluto}^2)^{1/3}, \quad (17)$$

where $\rho_{\rm atmo}$ is the density of Pluto's atmosphere at the surface, $\rho_{\rm imp} = 1000$ kg m$^{-3}$ is the impactor density, $H$ is the atmosphere scale height at the surface, and $R_{\rm Pluto} = 1189$ km is Pluto's radius. For Pluto's present-day atmosphere, where $\rho_{\rm atmo}$ is $10^{-4}$ kg m$^{-3}$, $H = 20$ km, and $T_u = 65$ K, these impactor size regimes are $r_{\rm min} = 0.13$ km, $r_{\rm cap} = 0.88$ km, and $r_{\rm gi} = 384$ km.

### 5.1. Impact Delivery and Erosion Model

In order to calculate the net delivery (or removal) of N$_2$ from impacts, we perform the following steps.

1. Generate an impactor with a randomly distributed radius $r$, following a size–frequency distribution proportional to $r^{-2.1}$.
2. Calculate the N$_2$ mass delivered by the impactor,

$$m_{\rm deliv} = \frac{4}{3}\pi r^3 \rho_{\rm imp} f_{\rm N2}, \quad (18)$$

where $f_{\rm N2}$ is the N$_2$ mass fraction of the impactor, assumed to be 2%. There is an implicit assumption that all of the N$_2$ is retained on Pluto's surface after the impact.
3. Calculate the atmospheric N$_2$ mass removed by the impactor, which is dependent on its size $r$ and also the parameters of Pluto's atmosphere at the given time, following Schlichting et al. (2015):

$$m_{\text{remov}} = \begin{cases} 0 & \text{if } r < r_{\min} \\ \dfrac{m_{\text{imp}} r_{\min}}{2r}\left(1 - \left(\dfrac{r_{\min}}{r}\right)^2\right) & \text{if } r_{\min} \leqslant r < r_{\text{cap}} \\ 2\pi \rho_{\text{atmo}} H^2 R_{\text{Pluto}} & \text{if } r_{\text{cap}} \leqslant r < r_{\text{gi}} \\ (0.4x + 1.4x^2 - 0.8x^3) M_{\text{atmo}} & \text{if } r \geqslant r_{\text{gi}} \end{cases}, \quad (19)$$

$$\text{where} \quad x = \frac{v_{\text{imp}} m_{\text{imp}}}{v_{\text{esc}} M_{\text{Pluto}}}, \quad (20)$$

where $v_{\text{imp}}$ and $v_{\text{esc}}$ are the impact velocity and Pluto's escape velocity and $m_{\text{imp}}$, $M_{\text{Pluto}}$, and $M_{\text{atmo}}$ are the impactor mass, Pluto's mass, and the mass of the atmosphere. We use an impactor velocity of $2 \text{ km s}^{-1}$ (Zahnle et al. 2003; Dell'Oro et al. 2013). The escape velocity from Pluto's surface can be calculated as $v_{\text{esc}} = \sqrt{2gR_{\text{Pluto}}} = 1.2 \text{ km s}^{-1}$. The mass of Pluto is $1.31 \times 10^{22}$ kg.

We then repeat this process of generating randomly sized impactors and track the amount of $N_2$ they deliver and remove until the total mass of the impactors reaches two ICRs. Given the random nature of the impactor sizes, we repeat this entire process 100 times and analyze the resulting distribution.

### 5.2. Impact Delivery and Erosion Results

According to Morbidelli et al. (2021), Pluto should have received two ICRs worth of impacting material in the past 4 Gyr, during which time Pluto was in its current orbit and its atmosphere was likely approximately equivalent to the present-day atmosphere. At 4 Ga, the Sun's bolometric luminosity was roughly 75% of the present-day value (Bressan et al. 2012), so Pluto's atmosphere would have been slightly cooler and more compact during this time period, but we assume the atmosphere was constant from 4 Ga to the present. Using Pluto's present-day atmospheric parameters ($\rho_{\text{atmo}} = 10^{-4} \text{ kg m}^{-3}$ and $H = 20 \text{ km}$), we estimate that two ICRs (equivalent to $5.8 \times 10^{17}$ kg of impacting material) will deliver $-74$ cm GEL, but they will also remove 10 cm GEL of atmosphere, resulting in a net delivery of $-64$ cm GEL of $N_2$. Note that, throughout this section, a negative sign means a net delivery of $N_2$, while a positive sign is reserved for a net removal or loss of $N_2$, to be consistent with the loss mechanisms discussed elsewhere in the paper.

Figure 8 shows the mass of $N_2$ delivered and removed by an impactor of a given size. From this, one can see that there is only a narrow size range in which an impactor is capable of removing more $N_2$ than it delivers. For an atmosphere like Pluto's present-day one, impactors between roughly 100 m and 2 km remove more atmospheric $N_2$ than they deliver. Impactors larger than 2 km, although significantly less likely to hit Pluto, deliver orders of magnitude more $N_2$ than they remove. The net effect is that impact delivery outstrips impact removal of $N_2$, resulting in a net delivery of $-29$ cm GEL per ICR, for a present-day Pluto-like atmosphere.

For the time period between the start of the giant planet instability and 4 Ga, Morbidelli et al. (2021) estimate that the Pluto system received two ICRs of impacting material as well, for a total of four ICRs between the start of the instability and the present day. This is a cumulative, integrated estimation, and the exact timing of when an impact of a given size would have occurred is not known. Thus, the atmospheric parameters at the time of each impact are also not known. Instead, we analyze three limiting cases of possible atmospheres during this time period: (1) the present-day atmosphere; (2) the coldest atmosphere experienced by the example Plutino, with $\rho_{\text{atmo}} = 10^{-6} \text{ kg m}^{-3}$, $H = 13 \text{ km}$, and $T_u = 27 \text{ K}$; and (3) the warmest atmosphere experienced by the example Plutino, with $\rho_{\text{atmo}} = 10^{-2} \text{ kg m}^{-3}$, $H = 60 \text{ km}$, and $T_u = 124 \text{ K}$. We assume that these atmosphere parameters are constant during the entire time period between the start of the giant planet instability and 4 Ga.

In scenario 1, Pluto experiences four ICRs of impacts with its present-day atmosphere. According to our model, these impacts deliver $-168$ cm GEL of $N_2$ and remove 22 cm GEL, for a net change of $-146$ cm GEL. A total of 30% of this amount occurs in the past 4 Gyr, with the remainder being delivered in the much shorter period of time between the start of the instability 4.537 Gyr ago and 4 Gyr ago.

In scenario 2, we assume that Pluto had a colder, more compact atmosphere between 4.537 and 4 Gyr ago and received two ICRs and then had its present-day atmosphere from 4 Gyr ago to the present and received two ICRs of impacts, for a total of four ICRs. We predict a net change of about $-155$ cm GEL of $N_2$, split into $-72$ cm GEL during the cold period and $-64$ cm GEL during the present-day atmosphere period. As shown by the blue curve in Figure 8, impactors greater than $\sim 0.3$ km radius will deliver more $N_2$ than they remove. In fact, two ICRs worth of impactors only remove a few cm GEL of atmospheric material. In the present-day atmosphere and in the cold, compact atmosphere, impact delivery always exceeds impact removal.

In scenario 3, in which we assume that Pluto had a very warm extended atmosphere between 4.537 and 4 Gyr ago, we predict a net change of $-58$ cm GEL $N_2$ delivered between 4.537 Gyr ago and the present. In the warm, extended atmosphere, it is relatively easier for impacts to remove more atmosphere than they deliver, compared to the colder atmospheres discussed above. It is possible for there to be a net removal (rather than a net delivery) for the time period immediately after the giant planet instability in the hot atmosphere case, with the median value being 26 cm GEL removed. However, over the full time range, the net effect is still a delivery of $N_2$ of $-58$ cm GEL.

Due to the higher rate of impacts in the aftermath of the giant planet instability, a significant fraction of the impact delivery and erosion occurs during the relatively short 100 Myr Wild Years. Estimates of volatile delivery or related quantities that use the present-day impact rate on Pluto and ignore the Wild Years time period will be significant underestimates as a result.

### 6. Atmospheric Jeans Escape

The final loss mechanism for $N_2$ investigated here is atmospheric escape via an enhanced Jeans escape process. At the time of the New Horizons flyby, Pluto's atmospheric structure implied an escape rate for $N_2$ of $3-8 \times 10^{22}$ molecules s$^{-1}$, while $CH_4$ escaped faster at $4-8 \times 10^{25}$ molecules s$^{-1}$ (Young et al. 2018). These escape rates were orders of magnitude lower than pre-flyby estimates.

Particles escape from a level of the atmosphere known as the exobase. Below the exobase, the atmosphere can be treated as a fluid, but above a transition region near the exobase, the atmosphere becomes collisionless, and the mean free path of a particle exceeds the scale height. This means that if a particle is traveling upward at the escape velocity from the body, then at the exobase, it is as likely to escape from the atmosphere as it is to collide with another particle. In the following section, we describe our scheme for estimating the atmospheric structure, including the height of the exobase, of Pluto's atmosphere during the Wild Years in order to estimate the escape rate.

### 6.1. Enhanced Jeans Escape Model

We use an enhanced Jeans escape framework (Zhu et al. 2014; Strobel 2021) as follows.

1. We begin by estimating the upper atmospheric temperature as a function of heliocentric distance and solar age. During the New Horizons flyby when Pluto was at 32.9 au, the upper atmosphere, above the $10^{-5}$ Pa level, was at $T_{upper} = 65$ K. To estimate how $T_{upper}$ varies with heliocentric distance, we create a scaling from Titan, which also has a hazy, nitrogen-dominated atmosphere. Using Titan's atmospheric temperature of 160 K at the $10^{-5}$ Pa level (Fulchignoni et al. 2005) and its semimajor axis of 9.5 au in addition to the values for Pluto, we derive a power law of $T_{upper} \propto h^{-0.725}$, where $h$ is the heliocentric distance.

2. The relation for $T_{upper}$ and heliocentric distance derived above needs to be adjusted to account for the variable output of the early Sun. Incident solar flux drops off as heliocentric distance to the inverse square power. This is equivalent to saying that the heliocentric distance is proportional to the incident flux to the $-1/2$ power. Thus, we can relate $T_{upper}$ to the incident solar flux as

$$T_{upper} \propto h^{-0.725} = (F_{bolo}^{-1/2})^{-0.725} = F_{bolo}^{0.3625}, \quad (21)$$

where $F_{bolo}$ is incident bolometric solar flux at a given heliocentric distance. In using the bolometric flux here, we are assuming that the atmospheric heating, namely, the haze heating and cooling, is dominated by visible-wavelength solar radiation. In the present-day Sun's output, the bolometric flux is 13,000 times higher than the UV flux (100–1180 Å) alone, so any UV haze heating is negligible. During the Wild Years, this drops to only 100 times higher, so UV heating could be nonnegligible. However, laboratory studies of Pluto's haze do not extend into UV wavelengths, so rather than make an unconstrained guess as to the haze absorption in UV wavelengths, we assume that haze UV heating and cooling are negligible even during the enhanced early Sun. Thus, we estimate $T_{upper}$ at any time and heliocentric distance as follows:

$$T_{upper}(t, h) = T_{upper,present} \left( \frac{F_{bolo}(t, h)}{F_{bolo}(t_{present}, 32.9 \text{ au})} \right)^{0.3625}, \quad (22)$$

where $F_{bolo}(t, h)$ is calculated using the PARSEC model described in Section 2.

3. The next step is to estimate the altitude $z_{upper}$ of the $10^{-5}$ Pa level. To do so, we assume that the atmosphere is isothermal and hydrostatic. During the New Horizons flyby, the atmosphere above the $10^{-5}$ Pa level was roughly isothermal at $T_{upper} = 65$ K due to efficient thermal conduction. Additionally, below the $10^{-5}$ Pa pressure level in Pluto's current atmosphere, it was proposed that haze radiative heating and cooling dominate the energy balance and control the temperature profile (Zhang et al. 2017; Wan et al. 2021). For simplicity, we assume the haze particles in the atmosphere also control the temperature below this pressure level throughout Pluto's history. We further assume that both haze heating and cooling rates roughly scale with the haze mixing ratio; thus, the temperature in the haze-dominant lower region is not very sensitive to the haze mixing ratio or the surface methane mixing ratio. Thus, in our escape model, we approximate the atmosphere as fully isothermal, from just above the surface through to the exobase. However, the gravitational acceleration is not constant with altitude (it varies by a factor of 2 or more between the surface and the upper atmosphere). Thus, the altitude $z$ of a given pressure level $P$ and temperature $T$ is

$$z = R_{Pluto} \left( \frac{k_B T}{m_{N2} g_{surf} R_{Pluto}} \log\left(\frac{P}{P_{surf}}\right) + 1 \right)^{-1} - R_{Pluto}, \quad (23)$$

where $g_{surf}$ and $P_{surf}$ are the gravitational acceleration and atmospheric pressure at the surface, $k_B$ is Boltzmann's constant, and $m_{N2}$ is the mass of a nitrogen molecule. For the atmospheric conditions at the time of the flyby, assuming an isothermal atmosphere leads to a prediction of 630 km altitude for $z_u$, compared with the observed altitude of 700 km for the $10^{-5}$ Pa level.

4. Given $P_{upper}$ and $T_{upper}$, we use the ideal gas law to calculate the particle density $N_{upper}$ at altitude $z_{upper}$. Then, we find the density profile at $z > z_{upper}$ assuming the atmosphere is isothermal above this altitude and again including an altitude-dependent gravitational acceleration $g(z_{upper})$:

$$N(z) = N_{upper} \exp\left( \frac{m_{N2} g(z_{upper})(R_{Pluto} + z_{upper})}{k_B T_{upper}} \frac{z - z_{upper}}{R_{Pluto} + z} \right),$$
$$z > z_{upper}. \quad (24)$$

Using this density, we calculate the mean free path of a nitrogen molecule,

$$l_{mfp}(z) = \frac{1}{N(z)\sigma}, \quad (25)$$

where $\sigma$ is the collisional cross section of a nitrogen molecule. We use $\sigma = 3 \times 10^{-19}$ m$^2$ (Strobel 2021). The exobase is classically defined as the altitude where the mean free path and the scale height are equal (Strobel 2021). The scale height is given by

$$H(z) = \frac{k_B T_{upper}}{m_{N2} g(z)}. \quad (26)$$

To find the exobase altitude $z_{exo}$, we set $l_{mfp}(z_{exo}) = H(z_{exo})$.

5. Next, we calculate the Jeans escape rate at the exobase,

$$F_{Jeans} = 4\pi (R_{Pluto} + z_{exo})^2 N(z_{exo}) \times \sqrt{\frac{k_B T_{upper}}{2\pi m_{N2}}} e^{-\lambda_{exo}}(1 + \lambda_{exo}), \quad (27)$$

where $\lambda_{exo}$ is the Jeans parameter at the exobase, given by

$$\lambda_{exo} = \frac{G M_{Pluto} m_{N2}}{k_B T_{upper}(R_{Pluto} + z_{exo})}, \quad (28)$$

where $G$ is the universal gravitational constant and $M_{Pluto} = 1.31 \times 10^{22}$ kg is the mass of Pluto. Recent numerical models have shown that molecules actually escape at a slightly different rate relative to the Jeans rate at the exobase. We adopt the enhancement model presented in Strobel (2021) based on the work of Zhu et al. (2014):

$$\Gamma(\lambda_{exo}) = e^{4.086446916 - 0.870018743\lambda_{exo}} + e^{0.63554463 - 0.019922313\lambda_{exo}}, \quad (29)$$

where we have corrected a small typo in the equation from Strobel (2021). For the range of Jeans parameters experienced by the Plutinos ($\lambda_{exo} \sim 9$–37), the enhancement factor $\Gamma$ ranges from 1.6 to 0.9. The net escape rate is then

$$F_{esc} = \Gamma(\lambda_{exo}) F_{Jeans}. \quad (30)$$

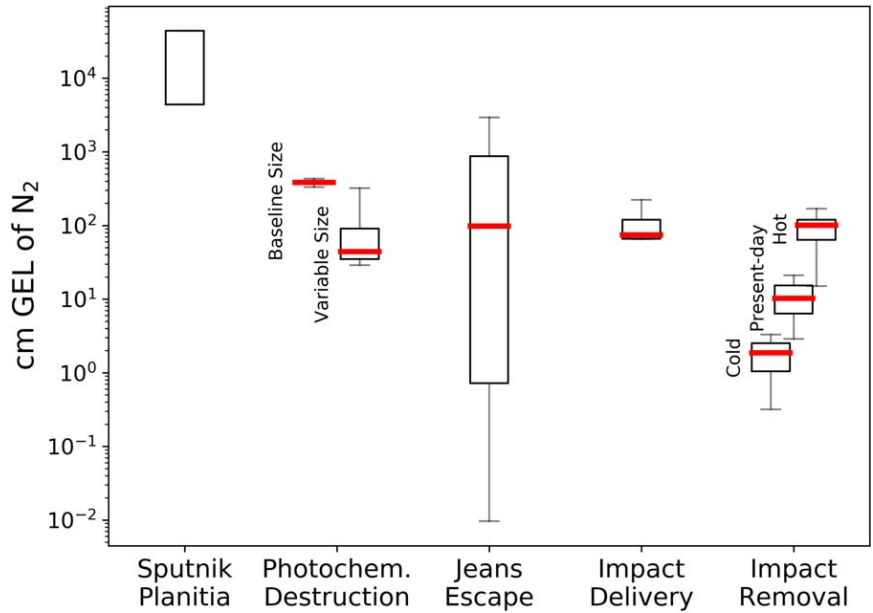

**Figure 9.** Whisker plot showing the amount of $N_2$ in Sputnik Planitia; the amount of $N_2$ lost during the Wild Years via photochemical destruction, atmospheric Jeans escape, and impact erosion; and the amount delivered by impacts. The red lines are the median values from the ensemble of 53 Plutinos for photochemical destruction and atmospheric escape and the 100 random samples drawn for the impact escape and delivery. The boxes encompass the 25th–75th percentiles, and the whiskers are the 5th and 95th percentiles.

### 6.1.1. Energy-limited Escape Model

Particles cannot escape at a rate faster than the incident energy driving their escape arrives at the atmosphere. The incident UV energy ranges from 0.85 to 0.2 mW m$^{-2}$ at 40 au from the start to the end of the Wild Years (Ribas et al. 2005). Conduction of heat up from the lower atmosphere into the exobase region is much smaller than the incident solar UV flux, on the order of $10^{-3}$ mW m$^{-2}$, assuming $\kappa(t) \approx 0.006$ W m$^{-1}$ K$^{-1}$ (Span et al. 2000) in the broad altitude range of decreasing temperature from 40 to 400 km. Thus, the maximum realistic escape rate is limited by the incident UV energy. To estimate this energy-limited escape rate, we use the model of Johnson et al. (2015). While it was developed to estimate escape rates from various KBOs by scaling from Pluto, we modify it to calculate escape rates from Pluto (or Plutinos) at various times throughout the history of the solar system. The relationship we use is as follows:

$$F_{esc}(h, t) = E_{UV+Ly\alpha} \left[ \left( \frac{32.9 \text{ au}}{h} \right)^2 + 0.09 \right] F_{esc,present}, \quad (31)$$

where $E_{UV+Ly\alpha}$ is the flux enhancement, accounting for both 100–1180 Å UV radiation and Ly$\alpha$ radiation; $h$ is the heliocentric distance in au; and $F_{esc,present}$ is the present-day energy-limited escape rate from Pluto, $2.6 \times 10^{27}$ $N_2$ s$^{-1}$ (Johnson et al. 2015). The 0.09 term in the brackets is due to the background flux from other stars.

For each Plutino migration from Nesvorný (2015), we know the orbital parameters at time steps of 10,000 yr. For each of these unique sets of orbital parameters, we calculate the heliocentric distance at 20 points per period, which is enough time steps to resolve the perihelion and aphelion. We then repeat this heliocentric distance as a function of time array for as many orbits as the Plutino completes in 10,000 yr. We assume that the Plutino restarts at perihelion each time the orbital parameters change. This technique gives us heliocentric distance as a function of time (albeit unevenly spaced in time) for the Plutino during its 100 Myr migration (the Wild Years). Using this and the solar evolution model, we can calculate the escape rate as a function of time or the total amount of $N_2$ that escapes during the Wild Years and throughout the entire 4.56 Gyr history of the solar system.

### 6.2. Atmospheric Jeans Escape Results

Figure 6(D) shows the estimated escape rate as a function of time for the example Plutino. At times when the calculated enhanced Jeans escape rate is greater than the energy-limited escape rate, we cap the escape rate at the energy-limited value (red line), although this rarely occurs. Integrating this capped escape rate over the length of the Wild Years yields a total nitrogen loss of 757 cm GEL for this Plutino, and the median loss of nitrogen from the ensemble of 53 Plutinos is 98 cm GEL (see Table 2). The range of escape losses is much larger than the range for the other loss mechanisms studied here. The middle 50% of the ensemble of Plutinos loses between 1 and 1000 cm GEL of $N_2$, as shown in Figure 9. The enhanced Jeans escape rate is highly dependent on the incident flux, so even a short period of time spent closer to the Sun than the average Plutino can have a large effect on the total $N_2$ loss.

In Figure 6, it is clear that at the end of the simulated time period, the escape rate is still far lower than the present-day rate. The solar luminosity is 71% of the present-day value at the end of the Wild Years, resulting in an atmosphere that is colder and more compact than Pluto's present-day atmosphere. The upper atmosphere temperature $T_u$ is 47 K, versus 65 K for present-day Pluto, and the exobase altitude $z_{exo}$ is 550 km, compared with 1700 km at present. As a result, the Jeans parameter (that is, the ratio of gravitational potential energy to thermal kinetic energy) at the exobase is 35 at the end of the Wild Years versus 16 for present-day Pluto, meaning that the atmosphere is much more tightly bound to Pluto and thus escape is much slower.

Using this same framework for estimating atmospheric escape, we calculate that the total nitrogen loss during the time from the end of the Wild Years to the present day is 2.35 cm GEL. Thus, the total nitrogen loss due to atmospheric Jeans escape is on the order of 100 cm GEL, which is a negligible amount relative to the observed nitrogen inventory on Pluto in the present day (i.e., the 44–440 m GEL of nitrogen contained within Sputnik Planitia). Thus, atmospheric escape via the enhanced Jeans regime is not a significant loss mechanism for nitrogen from Pluto at the present time or during the Wild Years.

Assuming instead that escape proceeded at the energy-limited rate throughout the entire simulation time or the age of the solar system gives an upper limit to the total amount of $N_2$ loss via atmospheric escape. During the Wild Years, we estimate that 1.2 km GEL would escape, using the energy-limited escape rate rather than the enhanced Jeans rate. Over the age of the solar system (including the Wild Years), the total amount lost would be 3.9 km GEL, which is about 10 times larger than the amount of $N_2$ ice contained within Sputnik Planitia. Only in this extreme case could atmospheric escape be a significant loss of $N_2$ from Pluto relative to the known present-day inventory.

## 7. Discussion

### 7.1. Comparison of Photochemical Destruction, Impact Erosion, and Jeans Escape

Sections 4 (Photochemical Destruction), 5 (Impact Erosion and Delivery), and 6 (Atmospheric Jeans Escape) are presented in order of decreasing significance to the nitrogen budget on Pluto. Photochemical destruction could remove up to 1270 cm GEL of $N_2$, the net effect from impacts could deliver up to −155 cm GEL of $N_2$, and atmospheric Jeans escape could remove up to 100 cm GEL over the age of the solar system. These values are summarized in Table 2 and shown graphically in Figure 9. During the Wild Years, photochemical destruction is the largest loss of $N_2$, followed by atmospheric escape and then impact erosion.

Sputnik Planitia, which likely contains the bulk of Pluto's present-day nitrogen inventory, amounts to 44–440 m GEL. In total, the three loss mechanisms analyzed here account for a net loss of $1270 - 155 + 100 = 1215$ cm GEL, or just over 12 m GEL. Thus, Pluto's primordial nitrogen inventory could have been 3%–30% higher than the present-day value.

### 7.2. Implications for Pluto's Composition

Glein & Waite (2018) also estimated Pluto's primordial $N_2$ inventory and concluded that unless Pluto's atmospheric escape rate was significantly higher in the past, the inventory would be dominated by the $N_2$ ice contained within Sputnik Planitia. In their "Large Loss" model, they assumed an energy-limited escape rate of $10^{27}$–$10^{28}$ $N_2$ s$^{-1}$ for Pluto, active over the age of the solar system (but they did not include any orbital migration; Pluto's orbital parameters were held constant at their present-day values), and calculated a net amount of $N_2$ lost via escape of 1.5–15 km GEL, which dwarfs the amount of $N_2$ in Sputnik Planitia (0.044–0.44 km GEL). In this scenario, Pluto's large $N_2$ inventory is most consistent with a solar-like composition. In their alternative scenario using the present-day escape rate over the age of the solar system ("Past Like Present"), Pluto's smaller $N_2$ inventory is consistent with a cometary composition. This means that the icy planetesimals that accreted together to form Pluto, which are assumed to have compositions similar to comets such as 67P/Churyumov–Gerasimenko (Rubin et al. 2015), could have delivered sufficient $N_2$ to explain Pluto's present-day inventory. In the "Large Loss" scenario, the comet explanation fails to deliver sufficient nitrogen to account for the observed amount of $N_2$ in Sputnik Planitia and the 1.5–15 km GEL lost via atmospheric escape.

The enhanced Jeans escape model that we use here predicts a slightly larger amount of $N_2$ lost to space than the "Past Like Present" scenario from Glein & Waite (2018), 100 cm GEL versus 8 cm GEL, but both are small relative to the amount in Sputnik Planitia. Our estimates of photochemical destruction and impact erosion/delivery are also small relative to the amount of $N_2$ contained within Sputnik Planitia. Thus, although this work uses a time-dependent escape rate based on the early Sun's output and incorporates Pluto's early orbital migration, we find that the loss of $N_2$ from Pluto over the age of the solar system is small. This is consistent with a cometary composition for Pluto, based on the work of Glein & Waite (2018).

Our escape analysis also has implications for the isotopic ratio of Pluto's nitrogen. Rayleigh fractionation in atmospheres occurs because molecules containing the heavier $^{15}$N atom are less likely to escape than those with $^{14}$N. However, a detectable change to the isotopic ratio $^{15}$N/$^{14}$N occurs only if substantial escape occurs, which we do not find here. Thus, Pluto's present-day isotopic ratio is likely indicative of the primordial isotopic ratio and therefore the isotopic ratio of the source of Pluto's nitrogen. New Horizons was not able to constrain the $^{15}$N/$^{14}$N ratio, but subsequent observations with the Atacama Large Millimeter/submillimeter Array have placed upper limits (Lellouch et al. 2017, 2022). Based on those upper limits, Glein (2023) concluded that $NH_3$ cannot be the sole primordial source of Pluto's $N_2$, and instead some initial $N_2$ and/or organic N is needed. Given our conclusion that escape was not significant (relative to the total inventory of $N_2$ on Pluto today), the "escape unimportant" case presented in Glein (2023) would apply (see their Figure 2(a)). While this case provides looser constraints than the case in which escape was important, it still constrains the initial source of Pluto's nitrogen to be at least 45% primordial $N_2$, organic N, or a mix of the two. If the isotope ratio $^{15}$N/$^{14}$N of Pluto's atmospheric or surface $N_2$ can be measured in the future, the primordial source and therefore the formation mechanism of Pluto could be determined with more certainty.

### 7.3. Consideration of Nonthermal Escape Processes

In Section 6, we discussed thermal escape from the atmosphere, in particular modified Jeans escape. However, there are other, nonthermal escape mechanisms that can contribute to atmospheric loss, which we will now briefly discuss. Escape process are categorized into five general groups: (1) thermal Jeans escape, (2) hydrodynamic escape, (3) photochemical escape (different but related to the photochemical destruction discussed in Section 4), (4) sputtering, and (5) ion escape.

Historically, Pluto's atmosphere was thought to be in the hydrodynamic escape regime (e.g., Trafton et al. 1997; Tian & Toon 2005; Strobel 2008), meaning that the escaping material can be treated as a fluid or bulk outflow moving at a single velocity. Models predicted $N_2$ escape rates of the order of $10^{28}\ N_2\ s^{-1}$ and were subject to the same energy-limit rate used in this work. They assumed an upper atmosphere temperature of 97 K for Pluto, but New Horizons observations indicate a much cooler atmosphere of only 65–68 K (Young et al. 2018), contributing to the lower estimation of Pluto's escape rate ($10^{22}\ N_2\ s^{-1}$) post-flyby. Post-flyby atmosphere models include cooling mechanisms such as water vapor (Strobel & Zhu 2017) or haze particles (Zhang et al. 2017; Lavvas et al. 2021; Wan et al. 2021). In our simple model here, we adopted the haze cooling scenario, resulting in a cooler temperature in the upper atmosphere and an enhanced Jeans regime for thermal escape instead of the hydrodynamic regime. Direct simulation Monte Carlo models have shown that Pluto's atmospheric escape remains in an enhanced Jeans regime and does not reach the hydrodynamic regime even with increased solar heating (Erwin et al. 2013; Zhu et al. 2014).

When $N_2$ is photochemically destroyed in the upper atmosphere, the constituent N atoms or ions can receive enough energy to escape directly. According to the photochemical model of Krasnopolsky (2020), most of the $N_2$ that gets destroyed combines with hydrocarbon radicals to form nitriles, predominantly HCN. Due to the strong triple bond, any HCN formed cannot be photochemically destroyed and thus precipitates to the surface (see Lavvas et al. 2021 for another, independent photochemical model of Pluto's atmosphere that includes deposition of HCN). The N atoms that do not get incorporated into other species can escape to space, and Krasnopolsky (2020) calculates that N will be lost to escape at a rate of $1.2 \times 10^{23}\ s^{-1}$, 50 times lower than the present-day $N_2$ photochemical destruction rate. We do not account for the direct escape of N during the Wild Years in this work.

Ion escape and sputtering both involve interactions between particles in Pluto's atmosphere and the solar wind. These interactions were not well studied prior to the New Horizons flyby, but Trafton et al. (1997) argued that nonradiative effects such as these could be significant at Pluto, due to its large distance from the Sun. However, New Horizons observations revealed that Pluto does not have strong interactions with the solar wind (Bagenal et al. 2016), at least at present. The early Sun had a stronger solar wind, with a mass-loss rate as much as 1000 times higher during the Wild Years relative to the present-day rate (Wood et al. 2002). Thus, during the Wild Years, sputtering may have been a significant loss mechanism of atmospheric volatiles, although quantifying the loss rate due to solar wind interactions is beyond the scope of this work.

### 7.4. Implications for Hypothesized Ancient Glaciation on Pluto

Pluto shows evidence for present-day glaciers and glacial processes. Sputnik Planitia is interpreted to be a large glacial ice sheet, and glacial flow is needed to maintain the level, crater-free surface that was observed by New Horizons (Bertrand et al. 2018). Additionally, flow features from the east of Sputnik Planitia that appear to empty into the basin are also interpreted to be active glaciation in the present day (Howard et al. 2017).

There is geologic evidence that implies that glacial processes were active in certain regions during ancient times as well. Howard et al. (2017) identify dissected terrains, containing a multitude of valleys with various morphologies, to the west, north, and northeast of Sputnik Planitia and favor an ancient glaciation interpretation for their formation. If the glaciers purportedly responsible for sculpting these valleys are sufficiently thick (1–4 km; McKinnon et al. 2016), they could exhibit basal melting, which increases the erosive ability of a glacier. Increased ice temperatures also would increase flow rates and reduce the thickness needed to induce basal melting.

Additionally, Howard et al. (2017) hypothesize that ancient glaciation could be responsible for dislodging the water-ice blocks from the bedrock surrounding Sputnik Planitia and emplacing them into the ice sheet itself, forming the observed mountain ranges, such as al-Idrisi Montes, near the edge of Sputnik Planitia. Sometime in the past, deep glacial flows from the surrounding highlands could have displaced the 10 km–sized blocks (perhaps created by fracturing of the ice crust during the impact that created Sputnik Planitia) and transported them to their present location 100 km from the edge of Sputnik Planitia. The timing of this transport is uncertain but could potentially be even more ancient than the glaciation that produced the highland valleys.

The loss mechanisms studied in this work, taken together, imply a total loss of $N_2$ on the order of 10 m GEL over the age of the solar system, which is mostly due to photochemical destruction of $N_2$. This potential loss of $N_2$ ice is far less than the kilometer thickness necessary to explain glacial carving of highland valleys or transport of water-ice blocks into Sputnik Planitia if the kilometer-thick ice mantle covered the entire globe of Pluto. If instead the kilometer-thick mantle of $N_2$ ice was restricted to only the latitude band where ancient glaciation is hypothesized ($-30°$S to $60°$N, $\approx 70\%$ of Pluto's surface area), the global layer of 10 m could explain a 14 m thick layer in this latitude band. If the $N_2$ ice mantle was restricted even further to only the localized regions of hypothesized ancient glaciation ($\approx 5\%$ of Pluto's surface), then a global loss of 10 m GEL would form a 200 m thick layer. Even in this very restricted region, the loss mechanisms analyzed here are insufficient to explain a kilometer-thick glacial layer, capable of carving valleys or transporting blocks. An alternative explanation would be stipulating that this ancient glacial activity occurred before essentially all of Pluto's volatile inventory became sequestered in the Sputnik Planitia basin and/or invoking a past climate epoch in which $N_2$ ice was restricted to the latitude band where the dissected valley terrains are observed. If escape proceeded at the energy-limited rate over the age of the solar system, Pluto could have lost close to 4 km GEL of $N_2$, which would be sufficient to explain the proposed ancient glaciation. As observed by New Horizons, the $N_2$ in Pluto's atmosphere is currently escaping at a rate 5 orders of magnitude lower than the energy-limited rate, so for this explanation to work, Pluto would need to be in a period of unusually low levels of escape at the time of the flyby.

### 8. Conclusions

The main conclusions of this work are summarized below.

1. During the Wild Years (30–130 Myr after the condensation of CAIs), Pluto likely spent <10 Myr closer to the

Sun than its present-day orbit and therefore <10 Myr at temperatures higher than the present-day 37 K. This likelihood is based on the median Plutino from our sample, which spent 9 Myr at heliocentric distances of less than 30 au. Typical heliocentric distances are between 15 and 80 au. Surface temperatures never exceed 63 K (the triple point for $N_2$, so liquid $N_2$ is never stable at the surface.)

2. The photochemical destruction rate of $N_2$ in Pluto's atmosphere was higher in the past due to the early Sun's enhanced UV flux, and we estimate a total loss of about 10 m GEL over the age of the solar system. This is the largest loss out of the three mechanisms studied here (photochemical destruction, impact delivery/erosion, and atmospheric escape). As much as 30% of the total photochemical destruction over the age of the solar system probably occurred during the short, 100 Myr Wild Years period due to the enhanced UV output from the young Sun.

3. Impacts are a net source of $N_2$ for Pluto, and impact erosion of the atmosphere is minimal in contrast. Impacts could have delivered up to −155 cm GEL of $N_2$ between the start of the giant planet instability and the present time. As much as 50% of the total impact delivery and removal likely occurred in the aftermath of the giant planet instability, when impact rates were higher than the present-day impact rate.

4. Atmospheric escape, calculated using an enhanced Jeans framework, is not a significant loss mechanism for nitrogen, during the Wild Years or otherwise. Over the age of the solar system, 100 cm GEL could have escaped, the large majority of which occurs during the Wild Years, when Pluto was closer to the Sun. If Pluto's atmosphere escaped at the energy-limited rate for the bulk of its history, then escape on the order of 4 km GEL is possible.

5. Sputnik Planitia alone contains 44–440 m GEL of $N_2$ ice, which dwarfs the potential loss due to photochemical destruction/escape or potential delivery by impacts. Thus, Pluto's primordial nitrogen inventory was likely very similar to its observed present-day inventory.

## Acknowledgments

This work was supported by NASA SSW grant NNX15AH35G. X.Z. was supported by NASA SSW grant 80NSSC19K0791. We would like to thank Randy Gladstone and Bill McKinnon for helpful conversations and their expert feedback on this work, as well as the participants of the New Horizons Summer Science Workshop in 2021 July for their early interest, encouragement, and advice for this work.

## ORCID iDs

Perianne E. Johnson  https://orcid.org/0000-0001-6255-8526
Leslie A. Young  https://orcid.org/0000-0002-7547-3967
David Nesvorný  https://orcid.org/0000-0002-4547-4301
Xi Zhang  https://orcid.org/0000-0002-8706-6963